# DECENTRALIZED AIR TRAFFIC MANAGEMENT FOR ADVANCED AIR MOBILITY

*Ítalo Romani de Oliveira, Euclides Carlos Pinto Neto, Thiago Toshio Matsumoto and Huafeng Yu*

*Boeing Research and Technology*

## Abstract

Leading proposals for Advanced Air Mobility (AAM) recognize the need for strategic and tactical airspace deconfliction, where the need for coordination appears in different forms and nuances. One recurring element is the use of pre-defined airways or corridors, a natural way to create order, with proven success from the conventional, manned, Air Traffic Management (ATM). But, while ATM is evolving to 4D Trajectory-Based Operations, when we apply the ATM principles to very dense and potentially more complex operations such as AAM, we have to consider their limitations in such demanding new environment. The requirement of following circulation corridors most often increase flight distance and inevitably create bottlenecks, hence we explore the hypothesis of not using corridors, testing such option via a simple and scalable simulation model.

Other motivations for comparing different forms of traffic coordination are redundancy and diversity, which have potential to increase system safety. Relying on a single method to maintain traffic separation of course would not be allowed in practice. However, the concepts that we have seen so far leave a gap between two very distinct and co-existing methods: one, cooperative and centered on a ground-based Provider of Services for UAM (PSU), and another, which is mostly non-cooperative and independent, centered on the individual aircraft, commonly referred to as Detect-And-Avoid (DAA). This duality achieves a welcomed diversity, however presents several points for improvement.

In-between these opposite methods, this paper analyzes the performance of an airborne cooperative method to coordinate traffic which is capable of safely solving conflicts of multiple aircraft (more than two) and achieve higher efficiency than DAA alone, thus with potential for being an alternative or a live fallback for ground-based traffic coordination.

## 1. Background

The current set up of intercity Air Traffic Management combines elements of airspace structures such as sectors, airways and terminal routes, with surveillance and communication means, and software for control and decision aid. These interconnected systems are highly mature and safe and are the most credible inspiration for Advanced Air Mobility / Advanced Aerial Mobility (AAM) [1] [2], although in this new environment the operations have new complexities if we consider the diversity of aircraft types, performances and obstacles.

Leading proposals [3] [4] make substantiated claims that the use of airspace structures in this environment can provide safety and efficiency benefits and therefore should be one of the fundaments in place. However, it is worth to observe that the obligation to follow circulation corridors decrease the efficiency of individual flights, a conclusion that is clear when no traffic conflicts are expected, but this loss may be compensated by the benefits of avoiding gridlocks and chaotic traffic. The corridors also create bottlenecks, although in the corridors they are more predictable and simpler to manage than if they occurred in random spots. By the use of fixed airspace structures, we are giving away some efficiency in order to decrease complexity and increase predictability, but we want to explore alternatives to achieve safe operations.

There are some studies comparing the capacity and efficiency of various degrees of structuration in the airspace, such as the study of [4], which shows that the unstructured airspace is the most efficient choice for fuel consumption, while a flexible way of layer structuration, although being slightly worse on efficiency, is slightly better on capacity, safety and stability, assuming that the aircraft are still manned, but with additional automation means for self-separation. One interesting aspect in this study is that higher degrees of structuration perform worse on almost all criteria, and from that we can deduce that airspace structuration should be used with moderation.

There are other elements which influence the safety and efficiency of operations, among which the conflict resolution and flow management algorithms. If there are shared routes and landing sites, for whatever airspace structure is chosen, conflicts are bound to occur and there must be means to resolve them. For intercity civil aviation, the conflict resolution is performed by hierarchically organized ATC where controllers act as arbitrators to solve conflicts among users, hence this is considered an organization centralized at the arbitrator. But there is a fundamentally different alternative approach for conflict management which is decentralized, where the aircraft resolve conflicts peer-to-peer or in larger self-organizing teams, without requiring a central arbitrator, but instead relying on appropriate protocols [5] [6] [7] [8] [9] [10] [11] [12] [13] [14] [15] [16].

The decentralized organization has the advantages of better workload distribution, less reliance on ground infrastructure and therefore fewer opportunities for existence of single points of failure (SPOFs).

## 2. Introducing the Simulation Models

One of the most attractive aspects of AAM is to provide transportation links between the densely populated or frequented city centers and the not-so-dense suburban areas, covering large territorial extents. At the dense areas, centralized traffic management can be reliable and resilient by the availability of high-speed networks and diverse means of redundancy, however that might not be the uniform on the suburbs. Besides, centralized algorithms tend to be more complex and they might have logical shortcomings which, theoretically, cannot be fully prevented not even by the most powerful verification techniques. Thus, the use of logically diverse solutions can contribute with additional reliability and safety. Therefore, we evaluate two scenarios where AAM employs decentralized conflict resolution and compare them with another scenario with centralized traffic management. We assume that, in all scenarios, all aircraft are cooperative.

### 2.1. Common Airspace Model

Simulation of full 4-dimensional (4D) maneuvering with multiple aircraft is mathematically complex and results in a control problem with infinite dimensions, so we employ a simpler model based on graphs. Its basic elements and definitions are in Figure 1 below.

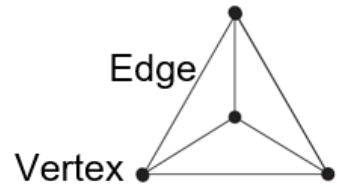

**Figure 1:** Basic elements of a graph used in the airspace model.

In this model, time is discrete and aircraft $i$ occupies vertex $v$ at time $t$, with
$$i \in [1, N]$$
$$v \in [1, V]$$
$$t \in [0, T-1]$$

Also, an aircraft $i$ occupies edge $(v_1, v_2)$ at time $t$.

A vertex corresponds to a vertiport or any sort of aerodrome used by the aerial vehicles, and the edges corresponds to the allowed routes between them. Of course, in order to have a minimum of realism in the model, we elaborate an airspace model with more instances of these elements, organized in a larger structure, as shown in Figure 2.

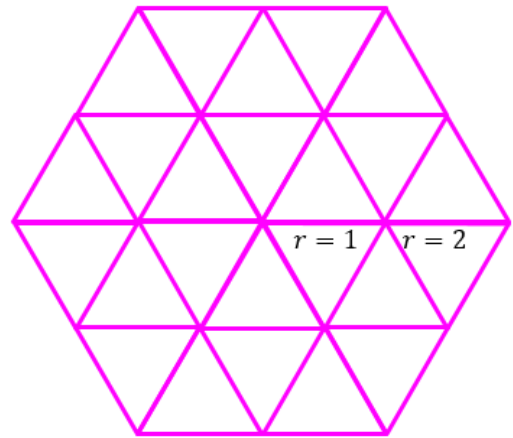

**Figure 2:** Airspace represented as a hexagonal lattice.

A graph composed as a regular array of a certain pattern can be called a *lattice*. In two dimensions, there are only 5 types of lattices, namely rhombic, parallelogramic, rectangular, square and hexagonal, this last one corresponding to Figure 2. In this case, the regular pattern is an equilateral triangle but, in despite of that, it can be called *hexagonal* because one single vertex can be connected with up to six neighbors, thus forming hexagons. Such hexagonal lattice has the following advantages over the other 2-D lattices:

- All angles between edges are the same, that is 60°. This property is also held by the rectangular and square lattices, in these cases with angle of 45°.
- All edges have the same size, a property shared with the rectangular and with specific cases of the parallelogramic lattices.
- Each vertex is connected to six other neighbors, instead of four for the other lattices. This allows more freedom to choose directions and hence allows more direct routes between vertices.

In this study, we restrict ourselves to 2-D or *planar* lattices, for the sake of simplicity. There certainly exist 3-D lattices that could be employed, but that would not change the fundamental concepts of traffic organization which are the focus of this study.

## 2.2. Implicitly coordinated conflict resolution

In this scenario, each aircraft avoids conflict based on common right-of-way rules and there is no explicit coordination of maneuvers, thus no need of data communication between the aircraft. This is also known as Detect-And-Avoid (DAA) [18]. The rules are designed so as that, when a conflict between two aircraft occurs, the evasive maneuvers of each aircraft have the correct symmetry and maintain safe separation between them.

In our airspace structure, there are only 5 possible angles of conflict, illustrated in Figure 3, where the ownship is at the bottom of the figure.

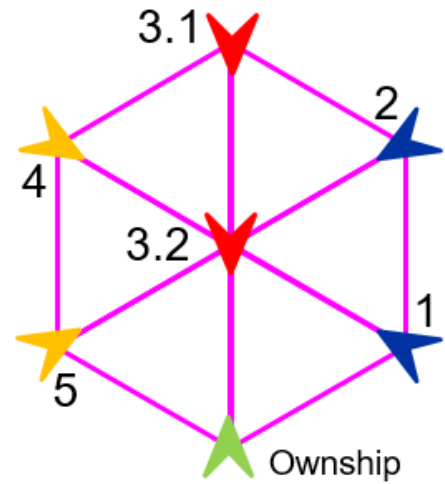

**Figure 3: Possible angles of conflict.**

This figure is relativistic, that is, it is applicable disregardfully to the heading angle of the ownship in relation to the so-called geographic north of the scenario. Our right-of-way rules are defined as:

Cases 1 (60°) and 2 (120°):
    Ownship turns right;
Cases 3.1 and 3.2 (180°):
    If, in the absolute geodesic referential, Ownship is heading towards the South hemisphere or exact West (180º):
        ownship turns right;
  else:
        ownship does not turn.
Cases 4 (240°), 5 (300°):
    Ownship does not turn.

Cases 3.1 and 3.2 differ only by the distance between the aircraft. Case 3.2 should happen only the aircraft were coming from different directions and one or both of them turn when there is only an edge separating them. This is the only case where the aircraft dispute the next edge to be used, instead of the next vertex, as in the other cases.

There is also the case where the aircraft are flying on the exact same line and direction, but in this case a conflict would happen only if the speed of the aircraft behind is higher than the speed of the aircraft ahead. In our case study, we rule out this case by imposing that all aircraft fly at the same speed.

It is easy to verify that these rules always keep a pair of aircraft separated in the lattice structure,

provided that the aircraft start at different vertices and, very important to notice, that there are no other aircraft (a 3$^{rd}$, 4$^{th}$, etc.) in the conflict. Instead, if there are yet other aircraft in the surroundings, there is no guarantee of conflict resolution. We may include additional rules to observe when there are more aircraft in the surroundings, however, the number of possible traffic configurations grows rapidly with the number of aircraft in the scenario and it is logically impossible to solve all situations without a higher level logic assigning unique priorities to the aircraft. Without that, there is always the possibility of entering deadlocks, livelocks or simply losing separation.

Establishing priorities [18] allows full resolution of conflicts, however to achieve that, complex logic is needed, and there is also the problem of who has the higher priorities, a problem which is not so easy to solve while at the same time maintaining equity among the aircraft. In our simulation model, we implement a compromise solution where the priorities of the aircraft are arbitrarily fixed in the beginning of the scenario; and use a non-exhaustive logic to determine the avoidance maneuver to be performed, which is not foolproof but goes beyond the requirements of DAA.

### 2.3. Strategic conflict resolution

Because of the limitations of implicitly coordinated conflict resolution, there is a considerable amount of work devoted to how to perform centralized conflict resolution for AAM [18] [19]. In some sense, this is not an entirely new development, since Air Traffic Control (ATC) represents the same concept for intercity traffic and it has been there for nearly a century.

In our simulation model for this scenario, we use the following assumption and features:
a) All aircraft will comply with whichever order is given by the ground control / PSU.
b) A Mixed-Integer Programming (MIP) algorithm is used to provide an optimized solution which:
   - Maintains separation (max. one aircraft per vertex / cell and edge at a time);
   - Makes that all aircraft go from origin to destination using the minimum collective amount of time (minimize the sum of individual flight times).
c) The conflict resolution determines the maneuvers for the entire duration of the scenario at once. In other words, the look-ahead of the conflict resolution algorithm has to be equal or larger than the time needed to all aircraft to fulfill their mission.

The last feature is very hard to achieve in reality, since the time spans can be considerably large (or even undetermined) until all the aircraft have stopped. Existing solutions of this type for Air Traffic Control simply truncate the operation time window and periodically recalculate the scenario with actual data, thus requiring a lower layer of continual tactical resolution to ensure separation and thus composing a hybrid solution. In our simulation model we can cover the entire operation window because both the number of flights and the flight durations are small.

In this setup, the most common interpretation is that the control decisions are made by the ground-based arbitrator (or controller) but, if this decision-making is entirely based on an algorithm, all aircraft could run the same algorithm in parallel to obtain the same result, which would be utopic but not impossible in controlled environments. In the case of a central arbitrator, it is obvious that a major failure or unavailability of its systems would stop or break the whole scenario; and, in the case of a common algorithm, a major flaw in it would have a similar effect. Considering the complexity of such algorithm, this is something to have in mind.

### 2.4. Collaborative conflict resolution

This scenario is in-between the previous two scenarios, in the sense that it does not require that every aircraft agrees on the same maneuvers to be performed by every aircraft, but requires that everyone agrees on the priorities for everyone, and that they explicitly coordinate the maneuvers with the nearby aircraft, following a certain airspace allocation protocol. This way, the conflict resolution algorithm does not need to cover the entire 4-dimensional scenario, thus being faster and less complex, allowing it to be executed individually by each aircraft and also providing more redundancy.

The so-called Collaborative Airspace Allocation protocol is described in [18] in detail and, besides having the features above described, it has also a key feature whereby near-term airspace resources (cells and edges) are allocated in an all-hands iterative "negotiation round" occurring at each periodic time step, and that aircraft turn or continue commitments are result of this negotiation round.

The key driver in proposing and using this alternative is to provide more redundancy to conflict resolution in mid-density and high-density airspaces, where implicit coordination falls short to solve conflicts with multiple aircraft, and strategic resolution require huge efforts in Verification & Validation and costly infrastructure investment.

## 3. Comparing the performances of the different conflict resolution techniques

Although none of the conflict resolution techniques above is meant to be used exclusively, it is worth to analyze their strengths and weaknesses in order to decide when and how to use each one.

Each of the three techniques are run in the same airspace model of Figure 2 with the same set of traffic configurations. The traffic configurations are distinguished by the number of aircraft (3 and 4) and are such that the aircraft start at the same time at different vertices with certain destinations that may be coincident or not. Also, only flight plans with the minimum length of 4 edges were allowed, in order to generate more conflicts and avoid wasting computing time in scenarios that naturally do not have conflicts. This way, we generated large number of traffic configurations with enough diversity so as to cover a large and balanced portion of the universe of traffic possibilities. This way, for the scenarios with 3 aircraft, there were 13,840 traffic configurations and, for the scenario with 4 aircraft, a total of 122,415. Higher numbers of aircraft were not tested because the computing time for 4 aircraft was already large.

As the model is bi-dimensional, only horizontal turn maneuvers are allowed, and those can occur only at the vertices. For the sake of simplicity, time is discrete and all speeds are equal (one edge per time step). To be consistent with the graph model, turns can occur only at the vertices. Air holds (stop and hover in the air) are not allowed in the cases here presented, because we made experiments with that feature and the results were worse.

Throughout the simulations, we collected some performance indicators, which will be summarized in the next sections.

### 3.1. Inefficiency

The inefficiency of a conflict resolution technique can be measured by the extra distance, time and fuel spent by the aircraft in consequence of the execution of resolution maneuvers. In our model, there is no representation of fuel and the time is linearly proportional to the distance, so it was enough to collect the flight distances per aircraft and divide by the optimum trajectory in unimpeded conditions (without other traffic). The mean results per scenario are shown in Figure 4.

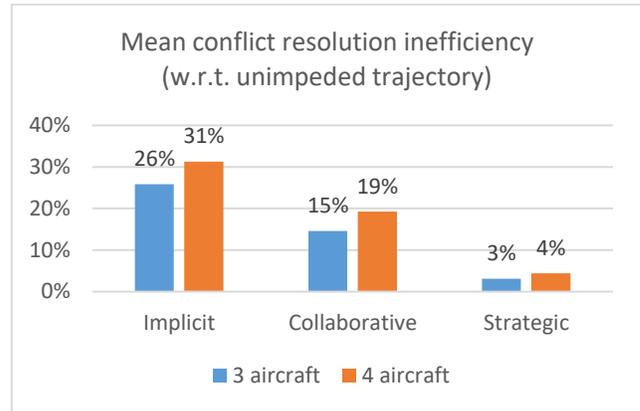

**Figure 4: Mean conflict resolution inefficiency.**

As expected, the Strategic resolution is the most efficient of them all, because it solves the traffic scenario globally and start-to-end. Implicit resolution is the most inefficient, because the logic to handle two or more simultaneous conflicts, or to handle second or higher degree conflicts, is highly relativistic and patchy. And, as expected, the Collaborative resulted in intermediate numbers.

### 3.2. Probability of fuel emergency

A particular type of highly inefficient scenario instance is the case where the chain of resolution maneuvers makes that the traffic return to the same collective state where they started prior to the resolution maneuvers and, in the absence of any perturbation to the aircraft, their logic will lead to an infinite number of repetitions of the same cycle, which we may call it as a *livelock*. Such occurrences are also a result of incomplete logic, whatever the causes of such incompleteness. And, instead of searching for the root causes of such incompleteness and trying to fix it, which may be mathematically impossible, we just use a simple criterion to identify such occurrences and stop the scenarios as if the aircraft were running out of fuel (or battery charge), thus we associate it to "fuel emergency".

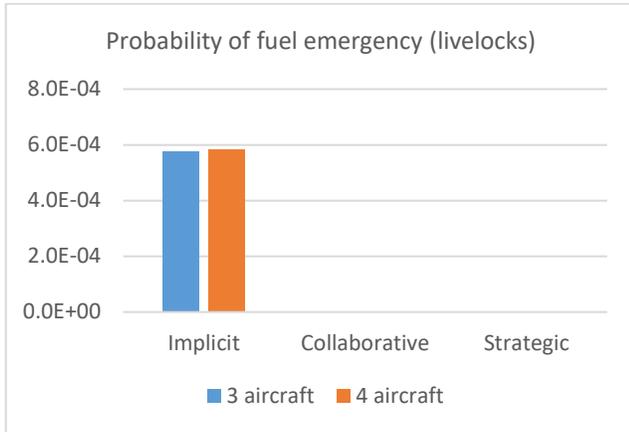

**Figure 5: Probability of fuel emergency.**

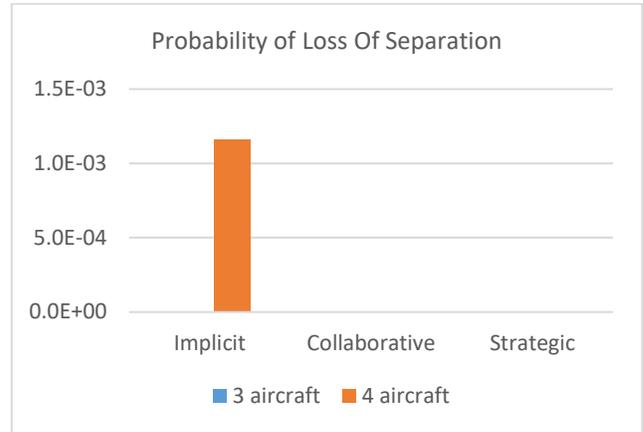

**Figure 6: Probability of loss of separation.**

With the number of occurrences of such cases, we calculate its probability of occurrence per each type of conflict resolution technique, as shown in Figure 5.

The fuel load used was 20 units of fuel, which allowed the aircraft to fly for 20 units of time or a distance of 20 edges in the graph. So, in Figure 5 we observe that only Implicit resolution presented such occurrences and, interestingly, the probability changes very little between the numbers of aircraft 3 and 4. These occurrences also contributed to the higher inefficiency observed in Figure 4.

### 3.3. Probability of loss of separation

In some other cases, the incompleteness of inconsistency of the conflict resolution logic results in not being able to keep separation among the aircraft. In our simplified airspace model, such occurrences necessarily imply in collision, but not in reality, where space and time are continuous and there are other factors to be considered. Hence we call this event as *loss of separation*, in the sense that aircraft become close to each other in space and time by a distance below a certain safety threshold. In our simulation model, this occurs when two or more aircraft occupy the same vertex or edge at the same time.

Throughout the scenario instances that we ran, the number of such loss of separation events were such to result in the probability shown in Figure 6.

As it can be seen, only Implicit resolution scenarios recorded such occurrences, but only when 4 aircraft are in the scenario, and not with 3. This fact demonstrates that our Implicit resolution logic is robust to avoid conflicts of 3 aircraft, in despite of the livelocks, and the other ones are robust to, at least, 4 aircraft.

### 3.4. Computing seconds per algorithm

Knowing the time needed to compute the resolution maneuvers is important because this gives an idea of how much of computing resources is needed to implement that solution in practice and, associated to this, that the required separation between the aircraft has to take into account the computing delay. For our three different algorithms, the mean computing times observed are in Figure 7.

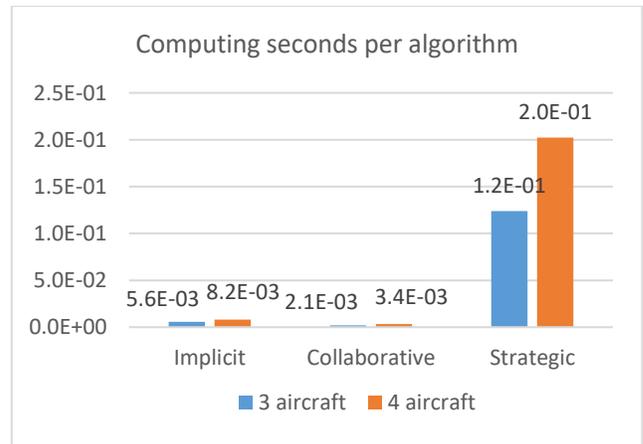

**Figure 7: Computing seconds per algorithm (excluding communications).**

The hardware used to run these simulations was a MacBook Pro with Intel i7 Quad Core at 2.2 GHz and 16 MB of RAM. The simulation programs were written in Python 3.8, and the Strategic resolution algorithm used `gurobipy` with an academic license as the MIP solver.

As expected, independently of the MIP solver, the Strategic algorithm requires a lot more of computing resources. It might be partially parallelized, depending on what the solver and the available hardware offers, but we still have not done research on this specific topic. The type of parallelization required here certainly rules out the distribution of computational load among the participant aircraft, due to the communication delays.

The other two algorithms performed much better because they are simpler in nature. Interestingly, the Collaborative one performed with less than half of the time required by the Implicit one, and some of the possible reasons for that are: Collaborative requires a look-ahead time of just one time step, while Implicit requires a look-ahead time of two time steps (needs some "estimation" of what will happen two steps ahead); and the livelocks generated by Implicit can be also a cause. Both algorithms are supposed to be ran in the embedded avionics, not on the ground, and if we separate the algorithm operations to be executed in each aircraft, the time per aircraft would be even smaller. However, it is important to point out that, in order to be fair, the total time for the Collaborative algorithm should include communication time among the aircraft, and for the Strategic the communication between the ground control center and the aircraft.

## 4. Equity considerations

Any conflict resolution algorithm requiring priorities among the aircraft is prone to inequity, that is, one aircraft having to deviate higher distances than others. Actually, this necessarily happens in single conflicts with aircraft having different priorities, and it may be unsafe to variate the priority of an aircraft in a single flight. This is the case of the Strategic and the Collaborative algorithms here presented. Thus, if we want to maintain equity among the aircraft, then we must recur to a higher level mechanism to update the aircraft priorities along successive flights, according to the accumulation of deviation distance incurred by each aircraft. We still have not done research on this topic but it is on our roadmap.

On the other hand, the Implicit algorithm is not inherently inequitable, however if the traffic configurations occur repeatedly with the same aircraft at each position and direction, then inequity will emerge from this repetition and there will be no possible mechanism to maintain equity, other than aircraft not flying the same trajectory at the same time each day.

## 5. Final Remarks

We presented a comparison of different conflict resolution algorithms using an extremely simplified model of airspace and aircraft dynamics. Although we believe that the use of more realistic models will not change the essence of the results, we would like to refine our algorithms to operate with better models and search for amplifications and attenuations of the effects here observed, when getting closer to reality. We also would like to take the steps necessary to perfect our algorithms and investigate how they can be used in practice, either as a stand-alone solution or hybridized with other algorithms.